# The reversal of the Sun's magnetic field in cycle 24

Alexander V. Mordvinov[1], Alexei A. Pevtsov[2], Luca Bertello[3], Gordon J.D. Petrie[3]

[1]Institute of Solar-Terrestrial Physics, Russian Academy of Sciences, Irkutsk, Russian Federation.

[2]National Solar Observatory, Sunspot, New Mexico 88349, USA.

[3]National Solar Observatory, Tucson, Arizona, USA.

**Abstract.**

Analysis of synoptic data from the Vector Stokes Magnetograph (VSM) of the Synoptic Optical Long-term Investigations of the Sun (SOLIS) and the NASA/NSO Spectromagnetograph (SPM) at the NSO/Kitt Peak Vacuum Telescope facility shows that the reversals of solar polar magnetic fields exhibit elements of a stochastic process, which may include the development of specific patterns of emerging magnetic flux, and the asymmetry in activity between northern and southern hemispheres. The presence of such irregularities makes the modeling and prediction of polar field reversals extremely hard if possible. In a classical model of solar activity cycle, the unipolar magnetic regions (UMRs) of predominantly following polarity fields are transported polewards due to meridional flows and diffusion. The UMRs gradually cancel out the polar magnetic field of the previous cycle, and re-build the polar field of opposite polarity setting the stage for the next cycle. We show, however, that this deterministic picture can be easily altered by the developing of a strong center of activity, or by the emergence of an extremely large active region, or by a 'strategically placed' coronal hole. We demonstrate that the activity occurring during the current cycle 24 may be the result of this randomness in the evolution of the solar surface magnetic field.

**Introduction**

The Babcock-Leighton mechanism [Babcock, 1961; Leighton, 1969] outlines a basic picture of cyclic changes of the Sun's magnetic fields. First, the solar cycle starts from a poloidal field defined by the magnetic field confined in the polar areas of the Sun. Then, differential rotation converts this poloidal field into a toroidal configuration, giving rise to the emergence of active regions in the photosphere. As the solar cycle progresses, the magnetic field of active regions is dispersed by turbulent convection and meridional flow. These transport mechanisms lead to the accumulation of magnetic flux of trailing polarity of decaying active regions at high solar latitudes, eventually reversing the polarity of the polar fields and building the next solar cycle. This concept led to the development of a distinct family of flux-transport numerical models that employ the Sun's differential rotation, supergranular diffusion and the meridional flows, to successfully represent many properties of observed long-term evolution of large-scale magnetic fields [DeVore et al., 1985; Wang et al., 1989].

The flux transport models were also extensively used to study the effects of various parameters on the evolution of polar magnetic field and the solar cycle [Jiang et al., 2013]. For example, Baumann et al. [2004] have shown that the active region tilt [described by Joy's law, Pevtsov et al., 2014], the diffusion and the rate of flux emergence have significant effect on the polar magnetic field. The speed of the meridional flow was found to affect the strength of solar cycle with slower meridional flow resulting in weaker solar cycles [Zhao et al., 2014]. Despite recent improvements, several questions remain open about flux transport models including the role of the not-well known variations in the speed of the meridional flow and scatter in orientation of active regions (Joy's law).

Both observations and numerical simulations indicate the importance of polar field as a predictor for strength of future solar cycle [e.g., Upton and Hathaway, 2014]. On the other hand, the observational evidence of the importance of active region tilts on the strength of the solar cycle is inconclusive. The initial report by Dasi-Espuig et al. [2010] about finding a relation between the mean active regions tilt of a given cycle and the strength of next cycle was questioned by Ivanov [2012] and McClintock and Norton [2013], and later, the results were revised by Dasi-Espuig et al. [2013]. Pevtsov et al. [2014] also argued that a relationship between the current surface activity (including active region tilt, flux emergence etc) and the strength of the polar field may be complicated by a prior state of the polar field. For example, a strong surface activity may not necessary lead to a stronger polar field if it has a significant polar field of opposite polarity to cancel out.

On the other hand, even a relatively modest surface activity may result in a strong polar field if the polar field of the previous cycle is weak. The question of the polar field strengths dependence on history is a long-standing issue in flux-transport research. For example, in their long-term flux-transport simulations Schrijver et al. [2002] found that the polar fields did not reverse during every cycle, e.g., a weak cycle would often fail to reverse strong polar fields. They suggested that this problem could be overcomed if the polar fields decayed away on timescales of 5-10 years. This idea of radial diffusion is not widely accepted now. Wang et al. [2002] showed that polar field reversals could be maintained if the surface flow speeds were systematically higher in large-amplitude cycles than in weak ones. Whatever this or other mechanisms can explain the complicated relationships between succeeding activity cycles of different amplitudes and polar field strengths is still the subject of debate.

The evolution of the polar magnetic field (and its reversal) in the current solar cycle 24 has been recently studied by several researchers [Mordvinov and Yazev, 2014; Sun et al., 2015; Petrie and Ettinger, 2015; Tlatov et al., 2015]. Still, a complete understanding of the processes affecting the recent polar field reversal is missing. This justifies additional studies of the peculiarities of current solar cycle and its polar field reversals. In our study, we use a combination of synoptic observations and numerical modeling as described in detail in Sections 2-5. Our findings are discussed in Section 6.

**2. Description of Data**

We use line-of-sight observations in the photospheric spectral lines of Fe I 630.15-630.25 nm taken by the vector spetromagnetograph (VSM) of the the Synoptic Optical Long-term Investigation of the Sun (SOLIS) facility [Keller et al., 2003; Balasubramaniam and Pevtsov, 2011] to investigate solar activity during the declining phase of cycle 23 and current cycle 24 (August 2003 - present). For early cycles we employ similar data taken by the NASA/NSO Spectromagnetograph (SPM) at the NSO/Kitt Peak Vacuum Telescope [Jones et al., 1992] from February 1974 to August 2003.

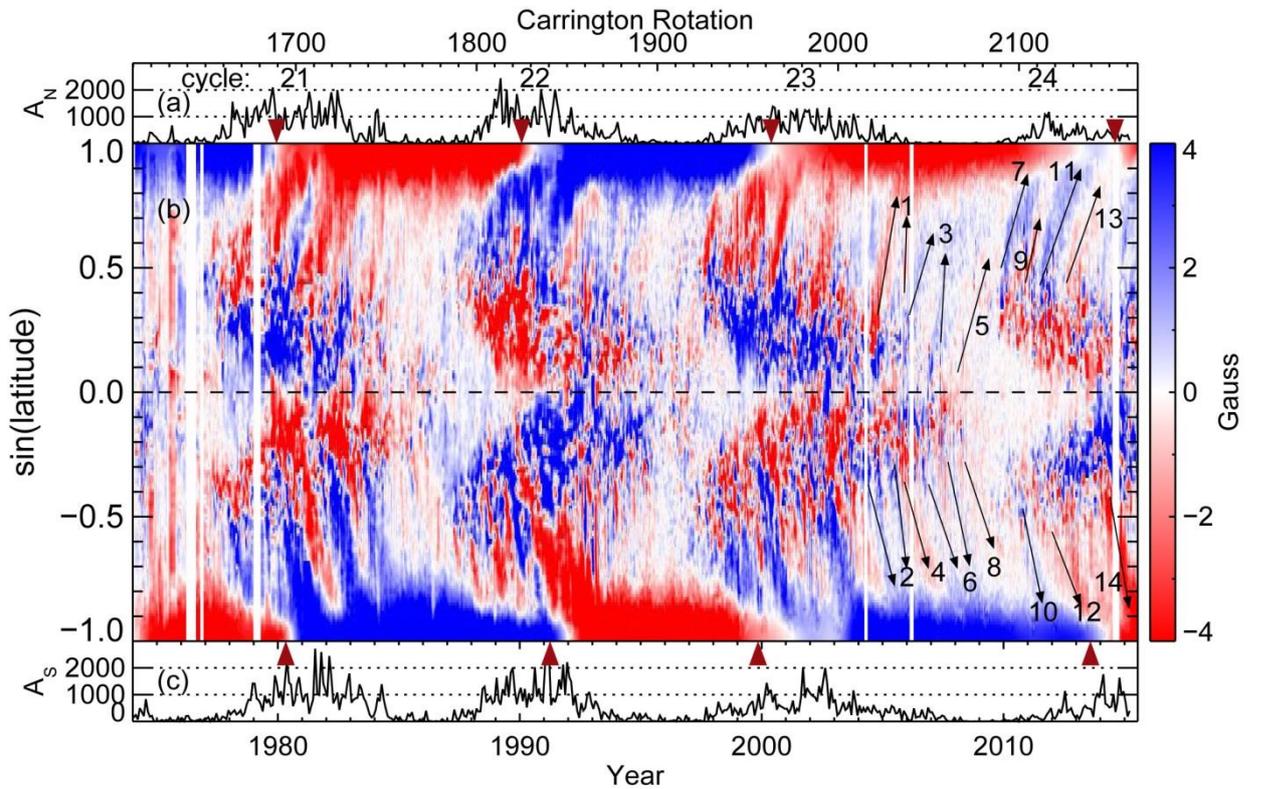

Figure 1. Changes in sunspot areas in millionths of solar hemisphere for the northern (a) and southern (c) hemispheres. Evolution of the Sun's magnetic fields from 1974 to 2015 (b). Blue/red halftones represent positive/negative polarity field. Numbered arrows mark position of poleward surges discussed in Section 4.

The butterfly diagrams shown in Figure 1, represents the long-term variations of large-scale magnetic field with time and latitude, is formed from the daily full-disk magnetograms as described in Petrie [2012]. Here we summarize the method. For each sky image, all pixels with central meridian distance 30º or less are binned into 180 equal-size bins in sine(latitude) and the average of each bin is taken. This process produces a 180-element array in sine(latitude) for each image. We combine these along a time axis to form the two-dimensional space-time maps shown in the figure.

In the construction of the butterfly diagrams, two corrections are applied: the longitudinal field measurements are used to derive data for the radial field component, and poorly observed and unobserved fields near the poles are estimated so that data are provided for all latitudes at all times. The radial field component is derived from the longitudinal measurements by assuming that the photospheric field is approximately radial, dividing by the cosine of the heliocentric angle ρ (the angle between the line of sight and the local solar radial vector).

Because the solar rotation axis is tilted at an angle of 7.25º with respect to the ecliptic plane, the fields near the solar poles are observed with very large viewing angles and are not observed at all for six months at a time. Also the noise level is inflated near the poles by the radial field correction described above. For these reasons, locations in the butterfly diagram nearest the poles are filled using estimated values for these fields. These estimates are based on a combination of direct field measurements, annual averages of high-latitude fields, and a polynomial fit across the pole for each image, calculated assuming symmetry about the pole. Finally the butterfly diagram is smoothed using a 27-day boxcar filter. White vertical stripes correspond to missing observations.

### 3. Activity at the end of cycle 23 and beginning of solar cycle 24

The current cycle 24 started after a deep and prolonged minimum, which ended in late 2008- early 2009 [Bertello et al., 2011]. During that minimum, the Sun's polar magnetic flux was significantly reduced, compared to the previous three cycles, and the background (non-polar) magnetic fields were very weak (Figure 1). There was no overlap between low-latitude activity of the preceding cycle 23 (see Figure 1, years 2006-2008) and higher latitude sunspot activity of cycle 24 (Figure 1, years 2010-2011). The earliest signs of cycle 24 activity can be identified at Carrington Rotation (CR) 2076, approximately at N40º latitude near the head of black arrow No. 5 in Figure 1.

From cycle 20 to cycle 23, Figure 1 shows a pattern of increasing gap between neighboring cycles as well as the increasing length of periods between polar field reversals in each hemisphere. For example, for the northern pole, the periods between polar field reversals were 3701 days (cycles 21-22), 3771 days (cycles 22-23), and 5191 days (cycles 23-24). For the southern pole, there were 3991 days (cycles 21-22), 3141 days (cycles 22-23), and 5021 days (cycles 23-24), accordingly. While at first glance it might appear that there is some correlation between the period of polar field reversals and the length of the following cycle, the statistics is very small; it could be just a simple consequence that the length of solar cycle together with the gap between cycles will approximately define the periods between polar field reversals. The small brown triangles mark the time of polar field reversals derived from WSO observations.

Figure 1 also demonstrates the uncertainty in determining the polar field reversals. Small brown triangles at the top and bottom edges of color panel mark the location of polar field reversals derived from Wilcox Solar Observatory (WSO) observations. While in some instances the location of triangles is in agreement with transition from one polarity field to the other as shown in the super synoptic map (e.g., Cycles 21 and 23, in the northern hemisphere, and cycle 21 in southern hemisphere), in other instances, the agreement is not very good (e.g., Cycles 22-24 in the southern hemisphere). Such disagreement may reflect a specifics of supermaps construction (e.g., averaging, or by the technique used to fill the polar field gaps) as well as the interpretation of direct observations of polar fields (e.g., smoothing the polar field measurements and estimating the polar fields for periods when either N or S poles are not observable from Earth).

The north-south asymmetry in magnetic activity can be identified in all cycles shown in Figure 1, but it is more pronounced in cycle 23. At the beginning of cycle 24, the hemispheric asymmetry may appear switching to favor the northern hemisphere (compare active regions activity in two hemispheres in 2011-2014), but later increase in sunspot activity in the southern hemisphere swung the asymmetry back to the southern hemisphere (Figure 1). The hemispheric asymmetry can be clearly seeing in the sunspot activity (Figure 1, panels a and b). For example, total area of sunspots in the northern hemisphere ($A_N$) peaks around year 2000, and it declines to its minimum in late 2006 (line plot at the top of Figure 1). By comparison, the total sunspot area in the southern hemisphere ($A_S$) reaches maximum in year 2002, and then it steady declines to its minimum late in 2008 (line plot at the bottom of Figure 1).

The solar cycle minimum lasts from about mid-2007 till late-2009 in the northern hemisphere, and from mid-2008 till early 2010 in the southern hemisphere. In cycle 24, the sunspot activity appears peaking in late-2011-2012 in the northern hemisphere, while sunspot activity in the southern hemisphere exhibits a peak in 2014. It is interesting to note that the cycle maxima in sunspot activity in two hemispheres were shifted by about two years in cycle 23 (with activity in southern hemisphere lasting longer). However, at the beginning of cycle 24, the phase shift between the two hemispheres was only about a few months (with the northern hemisphere leading in its activity). By the time of cycle 24 maximum, the difference in cycle maxima between the two hemispheres is again about two years, with activity in the southern hemisphere lasting longer.

Despite a relatively low amplitude in cycle 23, the asynchronicity in sunspot activity between two hemispheres helped in maintaining a relatively high level of magnetic activity during minimum of cycle 23. Muñoz-Jaramillo et al. [2015] have shown that although the minimum of cycle 23 was one of the lowest in recent history, it did not reach the lowest possible limit because the activity in each hemisphere reached minimum at different times. At the end of cycle 23, sunspot activity in the southern hemisphere lasted longer, as compared to the northern hemisphere.

In Figure 1b, inclined patterns show magnetic flux transport from decaying activity complexes towards the Sun's poles. Zones of intense sunspot activity resulted to extensive surges which reached the poles and led to the polar field reversals [Mordvinov and Yazev, 2014; Sun et al., 2015, Petrie and Ettinger, 2015]. Black arrows mark episodes of flux transport from decaying active regions during the declining phase of cycle 23 and rising phase of cycle 24. To emphasize the appearance of poleward surges, the magnetic flux is scaled between ± 4 Gauss. With such scaling an unphysical negative-polarity artifact appears prominent at the North Pole for 2015. It looks comparable in strength to the polar fields generally, but it is not. Our measurements indicate that the magnetic flux in this polar area continues to be close to zero, with mixed polarity but with a positive bias.

Between about CR2045 through CR2070 (or for about year and a half) sunspot activity was limited to the southern hemisphere only. This extended activity may helped the polar field in southern hemisphere to last longer. Thus, for example, in addition to poleward surges of negative polarity magnetic field (e.g., No. 4 and 8 in Figure 1) that were working to reduce the positive polarity flux in southern polar region, there were poleward surges of positive polarity (e.g., No. 2, 6, and 10) that continued strengthening the existing polar field of previous cycle. Similar behavior is observed in the northern hemisphere with surges of negative (No. 1, 5, and 9) and positive (No. 3, 7, 11) polarity field. Such sequential surges of positive-negative polarity are not uncommon (see Figure 1, years 1982-1986 southern hemisphere).

### 4. Origin of poleward surges of opposite polarity

Due to the preferred orientation of active regions relative to the equator (Joy's law), the following polarity magnetic field of dissipating active regions is transported poleward. According to Hale polarity rule, the active regions in northern hemisphere had negative (positive) trailing polarity in cycle 23 (24). The trailing polarity of active regions in the southern hemisphere was positive (negative) in cycle 23 (24). Thus, poleward surges No. 1, 7, and 11 (Figure 1b, northern hemisphere) and No. 2, 6, 12, and 14 (southern hemisphere) have polarity corresponding to normal Hale's and Joy's orientation of active regions in two cycles. Surges No. 3, 9, and 13 (Northern hemisphere) and No. 4, 8 and 10 (Southern hemisphere) have 'abnormal' polarity. To better understand the origin of each of these surges, we now provide a detailed description of their evolution.

#### 4.1. Surge No. 10

Early indications of surge No. 10 development can be traced back to NOAA AR11089. This region was located in the southern hemisphere and had the leading flux of positive polarity (see, Figure 2). Thus, the magnetic field of this active region was oriented in agreement with the Hale polarity rule for solar cycle 24. However, by CR2100 (August 21, 2010 at central meridian) the tilt of the magnetic field of this region did not follow the Joy's law. This non-Joy's tilt was the result of a very specific evolution of active region. The region emerged from behind East limb on July 19-20, 2010 as a complex multi-sunspot region. The overall orientation of active region was in agreement with Joy's law (leading polarity sunspots were located closer to the equator, and trailing polarity spots located father away from the equator, see Figure 2). As the region evolved, one of the sunspots of trailing polarity moved eastward from the main group, and dissipated rapidly. By July 23, this spot was reduced in size to a few small pores, which had completely vanished by mid-day July 24.

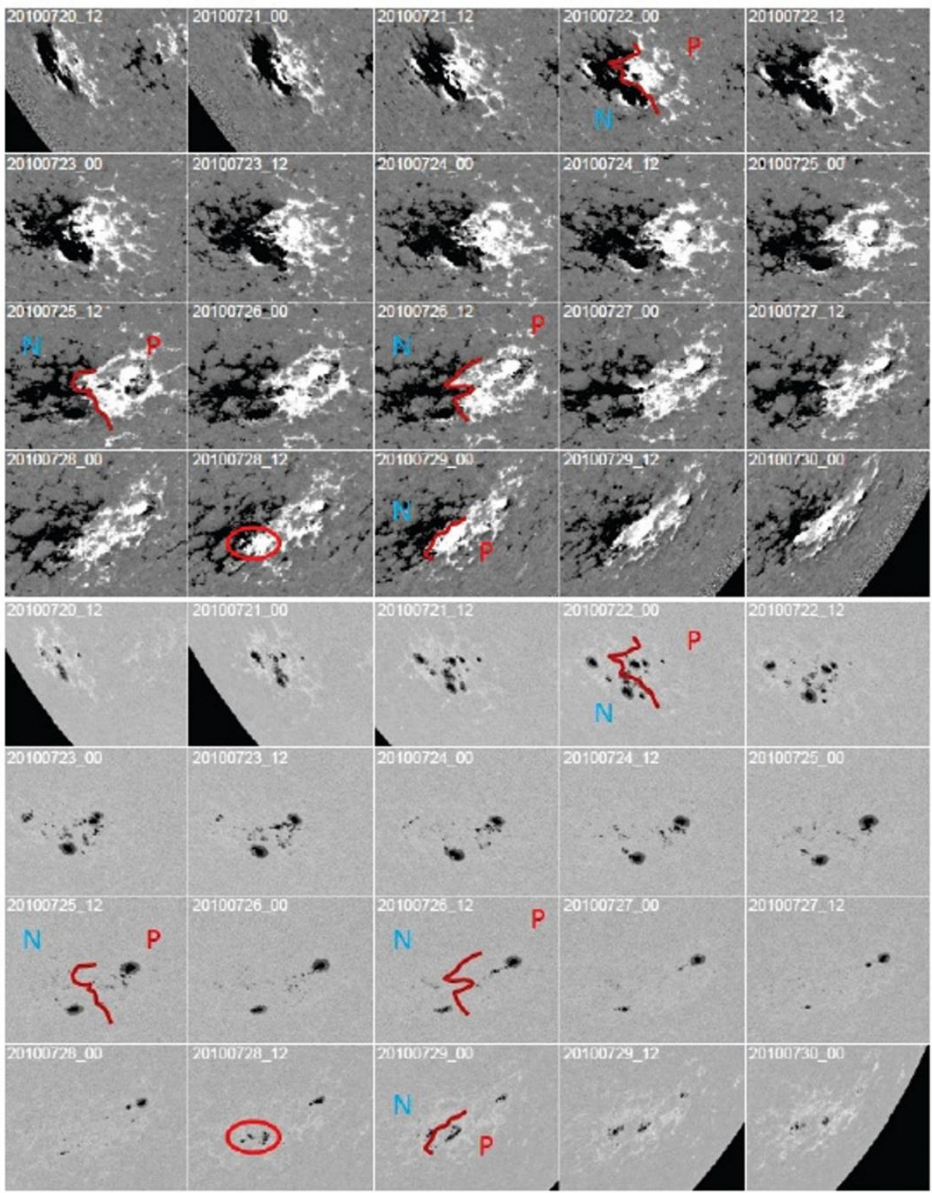

Figure 2. Evolution of active region NOAA 11089 as seen in line-of-sight magnetograms (top) and broadband pseudo-continuum images (bottom) from SDO/HMI. White/black halftones correspond to magnetic field of positive/negative polarity. Brown curve marks approximate location of magnetic neutral line. Letters P and N mark polarity (positive and negative) of leading and following parts of active region. Red oval outlines approximate location of new flux emergence at the tail of AR11089.

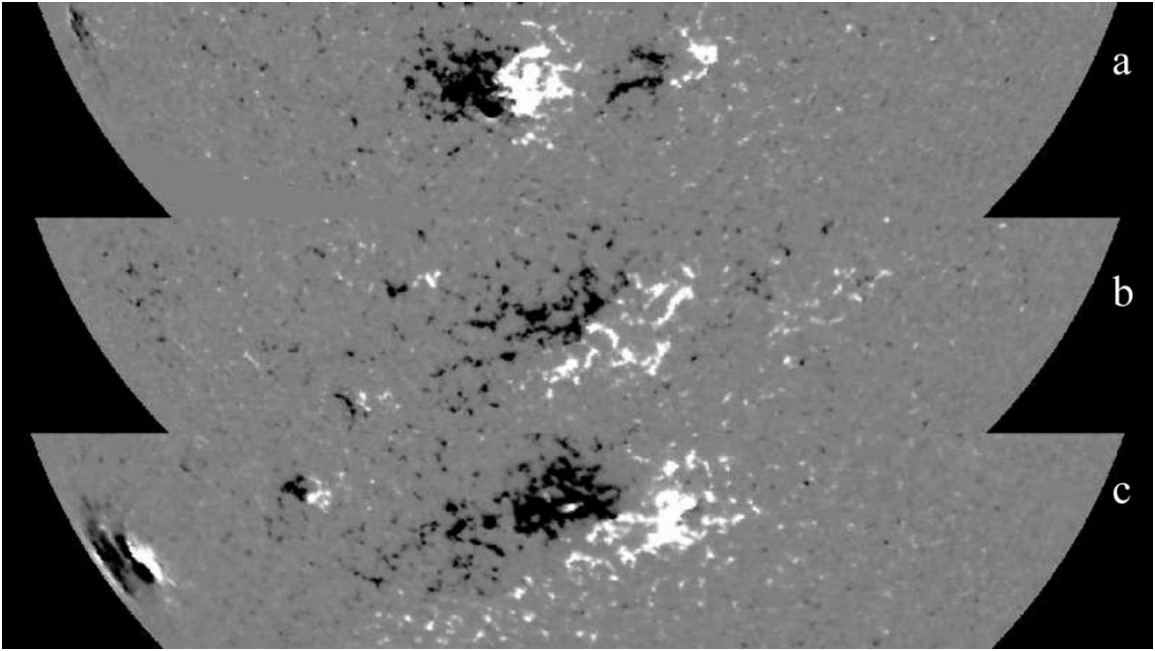

Figure 3. Evolution of magnetic ux at the source of poleward surge No. 10 over three solar rotations. Panel (a) show portion of full disk VSM/SOLIS longitudinal magnetogram taken 24 July 2010, 19:21:41 UT, (b) – 21 August 2010, 21:45:05 UT, and (c) – 17 September 2010, 16:26:30 UT.

The overall structure of active region got simplified, and by July 23, the active region could be best characterized as a typical bipolar region with two well-developed sunspots of opposite polarity and a few pores in between. The region's tilt was in agreement with Joy's law. After July 26, the sunspot of trailing polarity begins dissipating rapidly, and by early July 28, it is reduced to a few small pores. Then, in early July 28, a new bipolar region starts developing at the location of decaying flux of AR11089, and by July 29, this newly emerging flux appears as a bipolar group with small sunspots of leading and following polarity. The orientation of this region still follows the Joy's law, by the tilt is smaller then the tilt of AR11089. By the time the region reaches the West limb (July 30-31), its leading polarity field combined the positive flux from its original (but significantly dissipated) leading sunspot and the leading polarity of newly emerged region.

The trailing polarity is represented mostly by the flux of newly emerged region. As the result, the leading polarity flux is significantly larger in area as compared with the trailing flux. By its orientation, the magnetic flux now appears as having a non-Joy's tilt, although in white light the active region still follows the Joy's law in its orientation. One rotation later (CR2101), a new bipolar region developed at the exact location of dissipating remnants of AR11089. Leading and following polarity sunspots had a wide separation (about 15 degrees in longitude), and the region was oriented nearly parallel to solar equator (possibly a very slight non-Joy's orientation). Based on the appearance of magnetic flux on sequential Carrington rotation maps, this complex evolution placed the leading polarity flux to higher latitude and then the combined action of differential rotation and meridional flow led to formation of a 'tong' of dissipating flux of positive polarity gradually transported towards southern polar region. The next solar rotation (Figure 3b) shows dissipating remnants of active regions shown in Figure 2 developing a non-Joy's polarity orientation. The development of this surge continued as a new flux emerge in the same area further strengthening the non-Joy's orientation (Figure 3c).

In addition to NOAA AR11089, development of active regions NOAA 11108 (CR2101, Sept. 22 2010, Hale and Joy's orientation), NOAA 11115 (CR2102, Oct. 21, 2010, unipolar sunspot of positive polarity), and NOAA 11126 (CR2103, Nov. 17, 2010, non-Hale and non-Joy's orientation) also contributed to formation of poleward surge No. 10 although this contribution is less clear. ARs 11108, 11115, and 11126 developed at about the same latitudes ($\approx$ S30 deg), and their rotation rate was about 4-5% slower than typical rotation rate for this latitude. However, three regions developed in longitudes progressively shifted much farther eastward from previous rotation to consider them as part of the same center of activity (for example, a difference between location of AR11108 extrapolated to AR111026 time and actual location of this region is larger than 120 deg in longitude). The development of these three regions strengthens magnetic field of positive polarity in this general area, and this flux eventually contributes to tail of positive polarity that originated from NOAA AR10089.

### 4.2. Surge No. 13

This surge seems to develop starting from active region NOAA11417 (CR2120). There was a small coronal hole located south-west of that active region. On the next rotation, two new active regions (NOAA11432 and NOAA11433) developed east of decaying magnetic flux of AR11417. AR11432 was normal Hale's and Joy's region, but AR11433 was mostly unipolar (negative) polarity spot. Interaction between ARs 11432 and 11433 led to near dissipation of positive polarity field, and it strengthen the negative polarity flux. Coronal hole grew and extended to the west and north of these two ARs. Over the next several rotations, new active regions developed and dissipated to the east from this area, but the presence of the coronal hole helped retaining negative polarity field against dissipation. The emergence of active regions in this area was such that even development of a large active region 11476 in CR2123 did not change the balance of magnetic flux. It appears that the overall evolution led to positive field getting more fragmented and dispersed, while negative polarity field survived and contributed to the growth of area that still had coronal hole on some part of it. This evolution continues even after the coronal hole disappeared, and by CR2137-2139, a large area of negative polarity had extended to high latitudes, and it was later transported to the northern polar region.

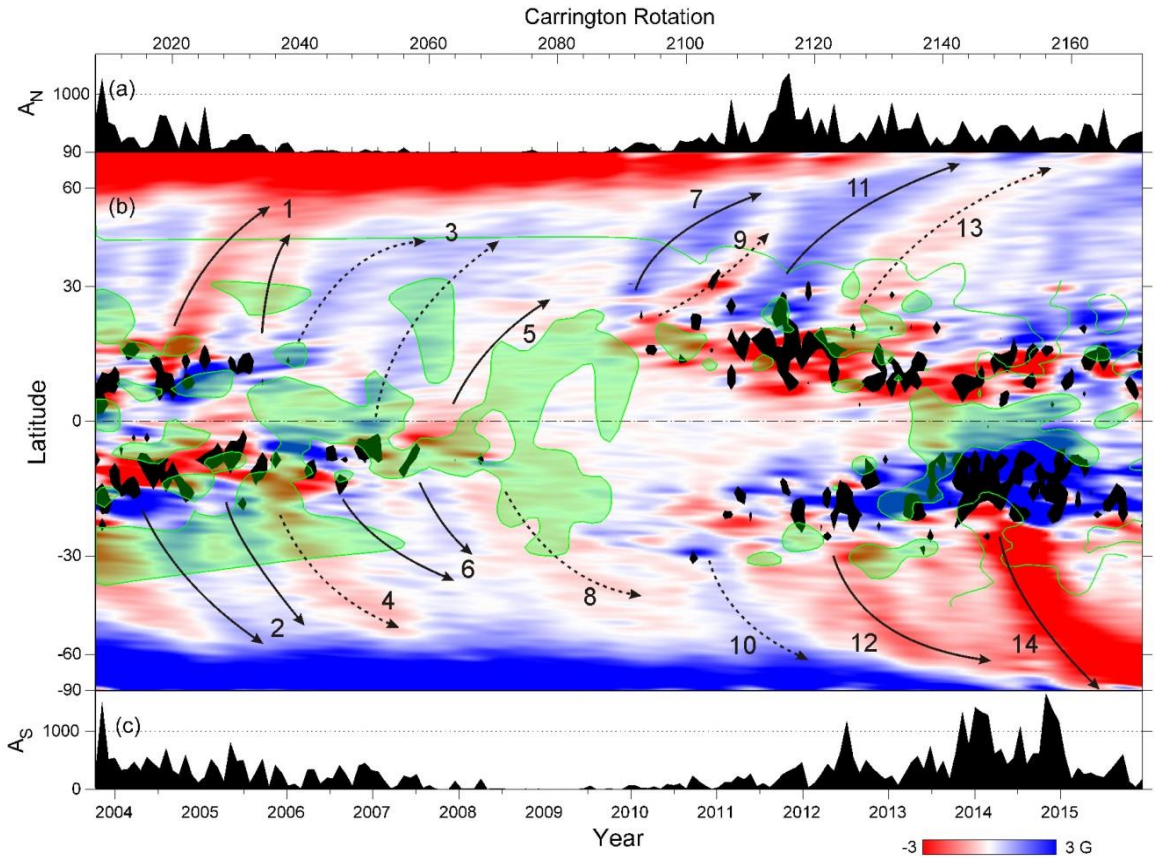

Figure 4. Changes in sunspot areas for the northern (a) and southern (c) hemispheres. Time-latitude evolution of zonal magnetic flux is shown in red-to-blue (b); zones of intense sunspot activity and domains of non-Joy's active region tilt are overplotted in black and green colors.

### 4.3. Surges No. 4, 3, and 9

Surge No. 9 originated in a large active region of non-Joy's orientation. The evolution of this surge was studied in detail (and modeled via a flux-transport modeling) by Yeates et al. [2015]. The reader is referred to this article for additional details about this surge.

The origin of the poleward surge 4, which starts around CR2138 cannot be traced to a single active region or activity complex. Instead, it appears to be the result of evolution of several active regions spread over a broad range of longitudes. Based on visual inspection of daily magnetograms taken during CR2137-38, it appears that the number of regions during this period of time show non-Joy's orientation in the southern hemisphere. As these regions evolve, the leading polarity field is transported to higher latitudes. Collectively, the surges from individual decaying regions contribute to what we see in supersynoptic map as poleward surge No. 4.

The origin of surge No. 3 is similar to surge No. 4: it also originated as the result of collective action of several active regions, and the solar activity that took place around the start-time of that surge also exhibits an enhancement the fraction of active regions with non-Joy's orientation. To verify the significance of active regions with non-Joy's orientation in development of poleward surges of opposite polarity, we plotted a distribution of active region tilts. Data for this plot were taken from the data set described by Györi et al. [2011].

To study evolution of zonal magnetic flux in more detail we analyzed synoptic maps composed of high-resolution SOLIS/VSM measurements. Figures 4a,c show changes in sunspot areas in the northern and southern hemispheres. Figure 4b shows a supersynoptic map composed of longitude averaged magnetic flux (in red-to-blue palette). Small gaps in these measurements were filled using SDO/HMI data. The zonal flux distribution is denoised using a wavelet decomposition technique. This diagram shows global rearrangements of solar magnetic flux in relation to sunspot activity in the current cycle. Zones of intense sunspot activity are shown in black (>70 millionths of the solar hemisphere). As a rule, surges of trailing polarities (marked with solid arrows) originated after decay of long-living activity complexes.

A location of active regions that disobey the Joy's law (colored in green) overplotted on the supersynoptic map of magnetic flux. It appears that that the active regions deviating from the Joy's law are not located randomly in latitude and time; there are large-scale patterns in their distribution. For example, in 2013-2014, majority of active regions in near-equatorial area had non-Joy's orientation (see large green area around years 2013 and 2014). Coherent areas of non-Joy's tilts also present in low latitude regions in late-2008- early 2009. The existence of such coherent areas contradicts a conventional explanation that non-Joy's tilts are the result of buffetings of magnetic flux tubes rising through the turbulent convection zone.

For most poleward surges discussed in Section 4, there is an area of non-Joy's tilt at the time and latitude when the surge had originated (see areas colored in green near the beginning of poleward surges No. 3, 9, 13 (northern hemisphere) and surges No. 4 and 6 (Southern hemisphere). Surge No. 10 does not show such association, which agrees with the description of evolution of active region tilt in that area. The presence of areas with non-Joy's tilt at the beginning of most surges indirectly supports the notion that majority of poleward surges of opposite polarity field originates from active regions with non-Joy's tilt. These surges are marked with dashed arrows.

### 4.4. Surge No. 5

This surge that originates in near equatorial region represents example when the magnetic field (of active region's leading polarity) crosses the equator, and is gradually transported to the polar areas of opposite hemisphere. The magnetic flux in this surge was quite weak, and it is not clear if this flux was eventually transported all the way to the northern polar region. This example suggests an interesting scenario that even if the sunspot activity is limited to a single hemisphere, the polar field in both hemispheres could still be created as it is the case for 'normal' cycles. The validity of such scenario depends on the interplay between the lifetime of decaying flux and the time it takes to transport it across equator to other hemisphere.

To further investigate this scenario for a cross-equatorial transport of magnetic flux, we employed a simplified flux-transport model previously used by us to study the effects of sunspot emergence and the dissipation on detectability of solar differential rotation from sun-as-a-star observations. For additional details about this model, see Bertello, Pevtsov, and Pietarila [2012]. In the model, we imitated the active regions using pairs of bipoles of different size. The bipoles were 'emerged' at random longitudes within the latitudinal range of about 5 degrees centered at 20 degrees of latitude in the southern hemisphere. All bipoles had fixed tilt relative to the equator, and their polarity orientation corresponded to the Hale-Nicholson polarity rule for the southern hemisphere in cycle 23.

The magnetic fluxes of bipoles were varied randomly within a preset range of fluxes. Test-runs of this model indicated that some of the magnetic flux of leading polarity can be transported across the solar equator as the result of active region growth (leading polarity is moving away from the center of emerging activity region in westward and the equatorward directions, while at the same time, the following polarity is moving slightly eastward and poleward). The diffusion of magnetic field also contributes to the cross-equatorial transport of magnetic flux. The simulations also showed that diffusion rate is important in defining if any flux will survive as a coherent structure by the time it is transported to the pole. Our findings from this model simulations are in agreement with Cameron et al. [2013] results.

Running the model over the period of time corresponding to about 5 years, we were able to see a build-up of magnetic flux of opposite polarity in the northern and the southern polar areas even though the sunspot activity was limited to the southern hemisphere. Switching the polarity orientation of bipoles (Hale-Nicholson polarity rule) from one cycle to the other, led to the reversal of polar fields in both hemispheres, as it would for normal cycles (with the sunspot activity in both hemispheres). As other example of transequatorial flux transport we refer the reader to Pevtsov and Abramenko [2010] who reported a case, when a coronal hole originating as an extension of a south polar coronal hole got disconnected and was gradually transported to the polar area in the northern hemisphere.

In the next section we study changes in cycle-integrated magnetic flux taking into account its prognostic importance. The polar field buildup quantifies the efficiency of magnetic flux transport and characterizes its north-south asymmetry in the current cycle.

**5. Cycle-integrated magnetic flux**

Solar and stellar dynamo are driven by convective and shear flows in solar convective zone. If the cycle-average properties of such flows do not change significantly from one cycle to the other, one could speculate that the total magnetic flux produced during each solar cycle should also show no significant variation from one cycle to the other. In other words, the total magnetic flux produced in cycle n, should be about the same as in cycle n-1 or cycle n+1. If such cycle-integrated flux-invariance existed, it would reveal itself in a correlation between the length of solar cycle and its amplitude, i.e., high-amplitude cycles would tend to be shorter in duration, while the opposite would be true for cycles with low amplitude. In fact, the international sunspot data do show a presence of such correlation albeit with a significant scatter [e.g., see Figure 5 in Petrovay, 2010]. Linear fit to the data corresponds to

$$A_{\text{cycle}} = (339.638 \pm 85.273) - (20.988 \mp 7.718) \cdot L_{\text{cycle}};$$

where $L_{\text{cycle}}$ is cycle length (years) and $A_{\text{cycle}}$ its amplitude (units of the international sunspot index). If one assumes that cycle 24 had reached its maximum, the fitted linear function can be used to estimate the length of cycle 24 as 12.3 years, which suggests that cycle 24 will reach its minimum in early 2021.

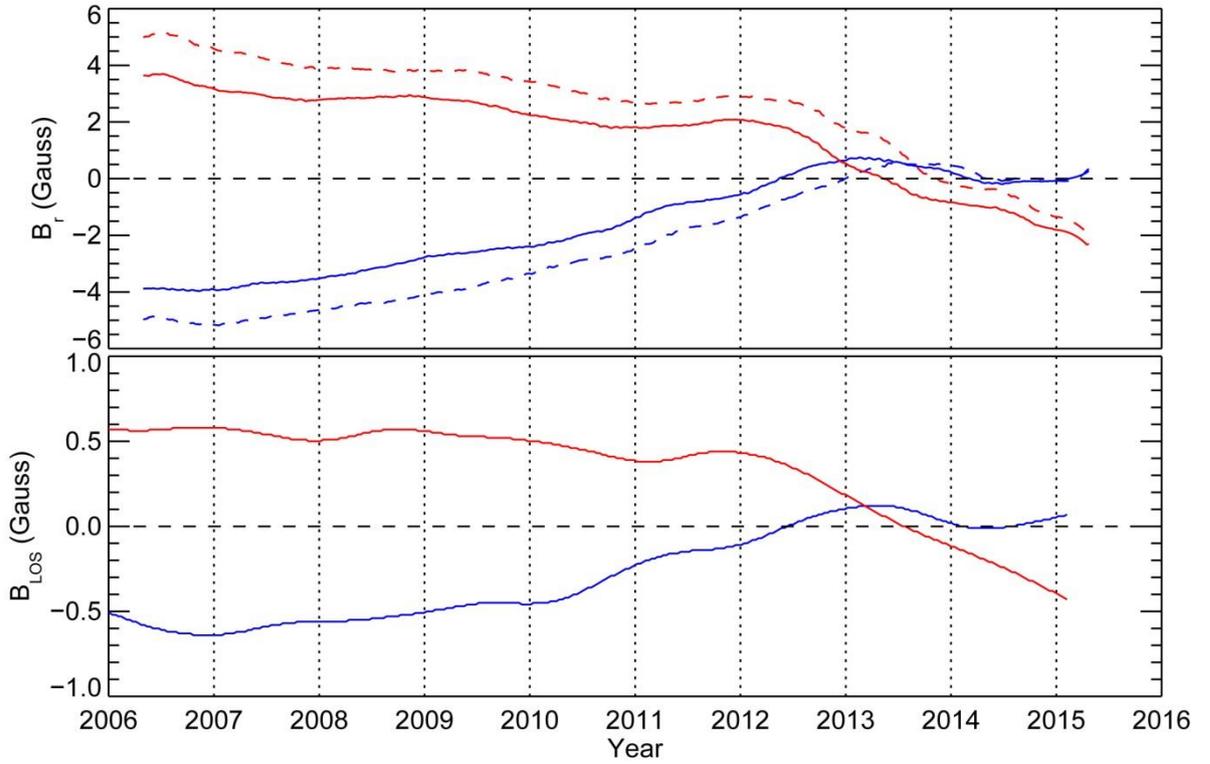

Figure 5. Comparison between SOLIS/VSM radial (top) and Wilcox LOS filtered polar measurements (bottom). Note the different scales. Northern hemisphere measurements are shown in blue, while the southern values are shown in red. For the case of SOLIS/VSM, measurements are shown for two different latitude bands: 60º to 70º (solid lines) and 65º to 75º (dashed lines). The times of polar reversals are given in Table 1. The most recent plot of polar fields can be found at solis.nso.edu.

Based on direct measurements of solar magnetic field by full disk longitudinal magnetographs operated by the National Solar Observatory (the 512 channel Diode Array Magnetograph (NSO-512) at the National Solar Observatory/Kitt Peak Vacuum Telescope (NSO/KPVT, 1974-1992), the NASA/NSO Spectromagnetograph at NSO/KPVT (SPM, 1992-2002), and the Vector Stokes Magnetograph on Synoptic Optical Long-term Investigations of the Sun (VSM/SOLIS, 2003-present), we find that the total flux observed in cycles 21, 22, and 23 is about the same. Difference in total cycle-integrated flux in cycles 21 and 22 is about -7.5% and in cycles 22 and 23 is about 1.7%. Slightly larger difference in pair of cycles 21-22 is due to lack of observations from the first two years of cycle 21. Thus, the direct magnetic field data also support the notion that the cycle-integrated magnetic flux on the Sun does not change significantly from one cycle to the other. However, while this apparent agreement in total cycle-integrated magnetic flux is intriguing, the speculation about the invariance of the cycle-integrated needs further confirmation, e.g., via direct dynamo modeling and analysis of other data sets such as WSO.

The integrated measurements of polar fields from the Wilcox Solar Observatory (WSO) and the SOLIS/VSM are shown in Figure 5. The data for this plot were computed following the recipe provided at the SOLIS web site at solis.nso.edu/0/vsm/vsm plrfield.html. The WSO data correspond to the line-of-sight flux, while the VSM magnetic field measurements were converted to radial flux (normal to the local solar surface) under the assumption that the photospheric field is vertical. This and a significant difference in the observational parameters between the two measurements can explain the difference in amplitudes of polar flux from these two instruments.

Table 1 provides approximate dates of polar field reversals derived from these two plots. Both WSO and VSM data were filtered using a low-bandpass filter to compensate for annual variations of observed polar field, which occur due to change in visibility of two poles with Earth orbital position. From these integrated data, it appears that the magnetic field in the southern hemisphere at 60º-70º latitudinal range reversed its polarity in mid-2013. At higher latitudes (65º-75º), the reversal occurred in late-2013.

The polarity reversal in the northern hemisphere is less clear: both WSO and VSM data indicate multiple reversals that occurred in mid-2012, mid-2014 and perhaps, late-2014-early 2015. Since the first reversal the polar field in the northern hemisphere remained weak, and it was practically fluctuating around zero. Only very recently (CR2161), with the increase in sunspot activity in the northern hemisphere we finally see a trend suggesting that the northern polar field is beginning to increase in amplitude (Figure 5, far right side of SOLIS/VSM plot). We see this behavior to be the result of a specific pattern of activity in the northern hemisphere.

### 6. Discussion and summary

The recent development of the solar activity in the current cycle made it evident that the North–South asymmetry of sunspot activity resulted in asynchronous reversal of the Sun's polar field. We demonstrate the well-defined surges of trailing polarities that reached the Sun's poles and led to the polar field reversals. We give concrete examples to demonstrate that the regular polar-field build-up was disturbed by surges of leading polarities which resulted from violations of Joy's law at lower latitudes.

During the declining phase of cycles 21 and 22, the dissipating field of active regions continued strengthening the polar field. In these cycles, the decline in polar field is clearly associated with a start of the next cycles. In cycle 23, however, the polar magnetic field in the northern hemisphere begun declining even before active regions of a new cycle 24 had emerged in high latitudes. This decline can be associated with poleward surge No. 3 (Figure 1), which originated from location of several small active regions with non-Joy's tilt. On the other hand, the polar field in the southern hemisphere got a slight boost, from surge No. 10. While this surge was eventually the result of non-Joy's orientation of corresponding magnetic flux, this abnormal orientation developed in course of a peculiar evolution of magnetic flux in this active region as well as additional flux emergence in this area in the following rotations.

In fact, Petrie and Ettinger [2015] found that poleward surges are almost always due to more than one region, particularly the important surges. Generally the high latitudes are occupied by decayed flux of both polarities from various regions at different stages of evolution. A surge develops when one polarity dominates overall, and sometimes, but not always, the dominant flux can be easily traced back to one, two, or a few major regions. After the first reversal, magnetic field in the Northern hemisphere had experienced a brief reversal to previous polarity state, which can be contributed to poleward surge No. 13. This surge has also developed as the result of a peculiar evolution and interaction between emerging active regions and long-lived coherent unipolar magnetic region which presence was outlined by a coronal hole.

Peculiarities appear rather random in their occurrence; we did not find any strong indication that this activity can be the result of some organized process. Nevertheless, a combined effect of this peculiar activity was sufficient to distort the process of polar field reversal in the northern hemisphere. The role of stochastic evolution in polar field reversal can be seen in other cycles. Thus, for example, comparing polar field evolution in cycles 21-22 with cycles 23-24, Petrie and Ettinger [2015] noted that the activity complexes were larger, longer-lived and tended to be arranged in a few giant structures at the height of cycles 21 and 22, whereas the cycle 23 and 24 active regions were smaller, less organized and shorter-lived.

The apparently more stochastic character of the cycle 24 reversal reflects the fact that it is the cumulative result of numerous relatively disorganized active regions. The elements of randomness in the polar field reversals described in this paper raise questions not only about how well our current modeling can predict the future solar cycle activity, but also if such prediction is even possible. Based on the success of the flux-transport models one can argue that the behavior of the solar cycle can be predicted sufficiently well, but then based on the examples shown in this article, these predictions may still be the subject of stochastic events that can significantly alter the course of the solar cycle. However, we also found that different cycles seem to produce about the same amount of cycle-integrated magnetic flux. We suggest that the latter could be used to estimate the length of cycle 24. Finally, the hemispheric asymmetry in sunspot activity may play a role in the strength of solar cycle. It appears that cycles with strong asymmetry tend to have a lower amplitude (e.g., cycles 23 and 24) in comparison with cycles in which sunspot activity in two hemispheres is more synchronized (e.g., cycles 21 and 22). Past suggestions that the sunspot activity during the Maunder minimum was restricted to only one solar hemisphere also support this notion. The role of the hemispheric asymmetry in strength of solar cycle will be the subject of future studies.

**Acknowledgments.** This work utilizes SOLIS data obtained by the NSO Integrated Synoptic Program (NISP), managed by the National Solar Observatory, which is operated by the Association of Universities for Research in Astronomy (AURA), Inc. under a cooperative agreement with the National Science Foundation. These data are freely available via the SOLIS web site at solis.nso.edu. AVM acknowledges support by the project II.16.3.1 under the Program of Fundamental Research of SB RAS. Data used in Figure 2 are courtesy of NASA/SDO and the HMI science teams.

**Table 1.** Time of polar reversals from Wilcox and VSM photospheric measurements.

|  | Wilcox 60º-70º | | SOLIS/VSM 65º-75º | | | |
| --- | --- | --- | --- | --- | --- | --- |
|  | North | South | North | South | North | South |
| 1st | 06/15/2012 | 06/26/2013 | 05/14/2012 | 05/10/2013 | 12/31/2012 | 10/28/2013 |
| 2nd | 03/03/2014 |  | 03/09/2014 |  | 05/19/2014 |  |
| 3rd | 10/15/2014 |  | 02/15/2015 |  | 03/13/2015 |  |